\documentclass[aps,twocolumn,PRB,reprint,superscriptaddress]{revtex4-1}
\usepackage{amsmath}
\usepackage{epsfig}
\usepackage{graphicx}
\usepackage{graphicx, color, epstopdf}
\usepackage{bm}
\usepackage{amssymb}
\usepackage{hyperref}
\usepackage{xcolor}
\usepackage{subfigure}
\begin{document}
\author{Min-Xue Yang}
\affiliation{National Laboratory of Solid State Microstructures and department of Physics, Nanjing University, Nanjing, 210093, China}
	
\author{Hao Geng}
\affiliation{National Laboratory of Solid State Microstructures and department of Physics, Nanjing University, Nanjing, 210093, China}
\affiliation{Collaborative Innovation Center of Advanced Microstructures, Nanjing University, Nanjing 210093, China}

\author{Wei Luo}
\affiliation{National Laboratory of Solid State Microstructures and department of Physics, Nanjing University, Nanjing, 210093, China}
\affiliation{School of Science, Jiangxi University of Science and Technology, Ganzhou 341000, China}

\author{Li Sheng}
\affiliation{National Laboratory of Solid State Microstructures and department of Physics, Nanjing University, Nanjing, 210093, China}
\affiliation{Collaborative Innovation Center of Advanced Microstructures, Nanjing University, Nanjing 210093, China}

\author{Wei Chen}
\email{Corresponding author: pchenweis@gmail.com}
\affiliation{National Laboratory of Solid State Microstructures and department of Physics, Nanjing University, Nanjing, 210093, China}
\affiliation{Collaborative Innovation Center of Advanced Microstructures, Nanjing University, Nanjing 210093, China}
	
\author{D. Y. Xing}
\affiliation{National Laboratory of Solid State Microstructures and department of Physics, Nanjing University, Nanjing, 210093, China}
\affiliation{Collaborative Innovation Center of Advanced Microstructures, Nanjing University, Nanjing 210093, China}
	
\title{Sign reversal of Magnetoresistivity in massive nodal-line semimetals due to Lifshitz transition of Fermi surface}
\begin{abstract}
Topological nodal-line semimetals offer an interesting research platform
to explore novel phenomena associated with its torus-shaped Fermi
surface. Here, we study magnetotransport in the massive nodal-line
semimetal with spin-orbit coupling and finite Berry curvature distribution which
exists in many candidates. The magnetic field leads to a deformation
of the Fermi torus through its coupling to the orbital magnetic
moment, which turns out to be the main scenario of the magnetoresistivity (MR)
induced by the Berry curvature effect.
We show that a small deformation of the Fermi surface yields a
positive MR $\propto B^2$, different from the negative MR by pure Berry curvature
effect in other topological systems. As the magnetic
field increases to a critical value,
a topological Lifshitz transition
of the Fermi surface can be induced,
and the MR inverts its sign at the same time.
The temperature dependence of the MR is investigated,
which shows a totally different behavior
before and after the Lifshitz
transition.
Our work uncovers a novel scenario of the MR induced
solely by the deformation of the Fermi
surface and establishes a relation between
the Fermi surface topology and the sign of the MR.
\end{abstract}
\maketitle
	
	\section{Introduction}
In a solid electronic system, the thermal and transport properties
are mainly determined by the electron distribution near the
Fermi surface \cite{kittel1996}.
Therefore, the change of Fermi surface by applying pressure,
doping or external field may lead to
considerable variation of the electronic properties
such as the specific heat and conductivity.
Interestingly, apart from continuous deformations,
the Fermi surface can also undergo abrupt
changes in its topology, so-called topological Lifshitz
transition \cite{lifshitz1960JETP}. At such a
critical point, anomalies in the
thermal and kinetic quantities may appear \cite{lifshitz1960JETP,nishimura2019prl},
which may even induce phase transition \cite{chu1970prb, norman2010prb,
sandeman2003prl,watlington1977prb,yamaji2007jpc,yelland2011np,liu2010np}.

Besides the topology of the Fermi surface geometry,
the electronic band structures can also have
nontrivial topology encoded
in the Bloch wave functions \cite{xiao2010rmp}.
In the past two decades, the study on such band topology has opened the
exciting research field of topological matters,
which includes topological insulators,
superconductors, and semimetals \cite{hasan10rmp,qi11rmp,armitage18rmp}.
The effects of the nontrivial band topology
are manifested mainly in two aspects:
the existence of exotic surface
states on the boundaries of the sample and
the modification of the dynamics
of the bulk electrons. The latter can give rise to
a variety of interesting transport phenomena
such as anomalous Hall effect \cite{nagaosa10rmp},
negative longitudinal magnetoresistivity (MR) owning to
the chiral anomaly \cite{armitage18rmp,wan11prb,murakami2007njp,son2013prb},
and most recently, the nonlinear Hall effect
induced by the Berry curvature dipole \cite{sodemann15prl,ma2019nat,du18prl,du2019nc}.

Given that the deformation of the Fermi surface and
nontrivial band topology may appear simultaneously,
it is interesting to explore possible novel effects
caused by the interplay between them. In particular, in
a topological material
with non-vanishing Berry curvature,
the interaction between the band topology
and the Fermi surface deformation
can be achieved by simply applying
an external magnetic field. The Berry curvature
results in a nonzero orbit magnetic moment (OMM)
of the electron \cite{chang1996prb,xiao2010rmp}, of which a coupling with
the magnetic field modifies the electronic dispersion
and accordingly the shape of the Fermi surface.
The benefit of this scenario is that the relevant physical
effects can be visibly revealed
by the MR measurement.
For example, it was shown recently that
the coupling of the OMM
to the magnetic field in the Weyl semimetal
induces a deformation of the Fermi surface from a sphere
to an egg shape \cite{knoll2020prb}.
As a result, the negative MR
due to the chiral anomaly can even have the opposite sign
as the inter-node scattering is
taken into account simultaneously \cite{knoll2020prb}.
Besides the smooth deformation of the Fermi surface,
topological Lifshitz transition of the Fermi surface
is studied as well in various topological materials \cite{volovik2017ltp,zhang2017nc,
volovik2018,yang2019nc,ekahana2020prb,wu2015prl}, which is driven by
different methods such as the temperature. An interesting question
that arises is whether both the smooth deformation
and the abrupt Lifshitz transition of the Fermi surface
can be implemented by imposing a magnetic field, such that
the underlying physics can be manifested under the same framework
of MR measurements. Apparently,
systems with a Fermi sphere
of zero genus is hard to achieve this purpose.

\begin{figure}
\centering
\includegraphics[width=0.45\textwidth]{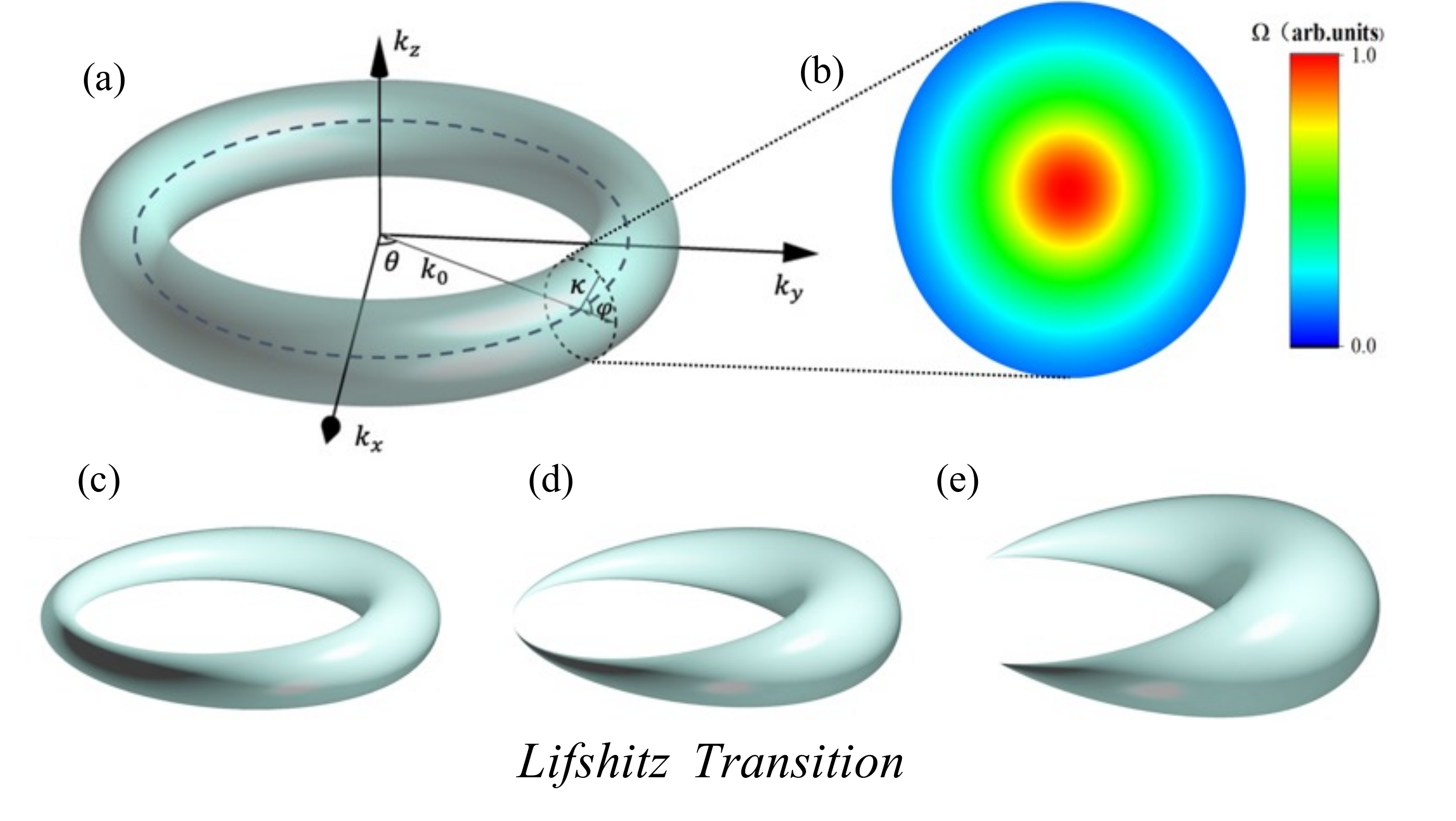}
\caption{(a) Schematic diagram
of the torus-shaped Fermi surface in nodal-line semimetal,
with the major radius $k_{0}$, minor radius $\kappa$, toroidal
angle $\theta$, and poloidal angle $\varphi$. (b) Finite
distribution of Berry curvature in the poloidal cross section. (c)-(e)
Fermi surface deformation induced by the magnetic field,
with (c) a slight deformation, (d) critical point of
topological Lifshitz transition
and (e) Fermi surface with genus one.}
		\label{fig1}
	\end{figure}

In this work, we study the topological Lifshitz transition
of a torus-shaped Fermi surface with genus one
in a doped massive nodal-line semimetal [cf. Fig. \ref{fig1}(a)] driven by a magnetic field
and predict the corresponding transport signatures.
Without a gap opening,
the nodal-line semimetals are characterized
by the linear band crossing along open lines or closed loops
which have been confirmed in a variety of candidates
\cite{Burkov11prb,Kim15prl,Yu15prl,Heikkila11jetp,Weng15prb,Fang15prb,
Chan16prb,Zhao16prb,Bzdusek16nat,Chenwei17prb,
Yan17prb,Ezawa17prb,bian2016nc,schoop2016nc,neupane2016prb,
topp2016njp,takane2016prb,hu2016prl,hu2017prb,kumar2017prb,pan2018sr,li2018prl,zhang2018fop,li2018prb,chen2020prb,an2019prb,zhou2020jcp}
and may exhibit interesting transport
properties \cite{chen2018prl,chen2019prl2,Luo21scpma}.
The topological property of the bands
is encoded in the $\pi$ Berry flux carried by the nodal line,
while the Berry curvature vanishes elsewhere in the Brillouin zone.
As a result, the OMM and its coupling to the magnetic field are absent
without spin-orbit coupling.
However, sizable spin-orbit coupling generally exists in
many candidates of the nodal-line semimetals, which may
give rise to a gap $\Delta$ of order 10-100 meV in the energy
spectrum \cite{zhang2019jpcc,chen2019prl}. In these materials,
the $\pi$ Berry flux
spreads into a finite-region
distribution of the Berry curvature
in the momentum space; see Fig. \ref{fig1}(b).
As the sample is slightly doped,
a deformation of the Fermi torus
can be achieved by coupling the OMM with
the external magnetic field $B$; see
Figs. \ref{fig1}(c)-\ref{fig1}(e).

Here, we study the magnetotransport in the massive
nodal-line semimetals in the diffusive limit
by using the semiclassical Boltzmann formalism
and taking into account the Berry curvature effect.
We show that
the OMM induced deformation of the Fermi surface
is the key scenario for the MR, in contrast to
the negative MR in other topological materials
by pure Berry curvature effect without OMM \cite{son2013prb,dai2017prl}.
Specifically, it is found that a slight deformation of the Fermi torus
results in a positive MR $\propto B^2\Delta^2$;
Remarkably, as $B$ increases to a critical value,
the topological Lifshitz transition occurs
for the Fermi surface [cf. Fig. \ref{fig1}(d)], which is proved to
coincide with a sign reversal of the MR.
Our results provide a new perspective to understand the
MR in nodal-line semimetals and
establish a fascinating relationship
between the sign of MR and the Fermi surface topology.

The rest of this paper is organized as follows.
In Sec. \ref{II}, we calculate the OMM in the massive
nodal-line semimetal and its coupling to the magnetic field;
In Sec. \ref{III}, we derive the general conductance formula
using the semiclassical Boltzmann equation.
In Sec. \ref{IV}, we study the magnetic
transport properties and their physical origins
both analytically and numerically. In Sec. \ref{V},
we give a discussion on the
different manifestation of our scenario
from others.
In Sec. \ref{VI}, we draw our main conclusion.

\section{OMM in massive nodal-line semimetals}\label{II}
We start with a minimal model of the nodal-line semimetal
with two bands crossing along a single nodal loop as \cite{Burkov11prb}
\begin{equation}\label{1}
H_0=\hbar \lambda (k_{x}^{2}+k_{y}^{2}-k_{0}^{2}) \tau_{x}+\hbar v k_{z} \tau_{y},
\end{equation}
where $\tau_{x,y}$ are the Pauli matrices acting on
the orbital degree of freedom.
The eigenvalues of the Hamiltonian $\epsilon_{\bm{k}}=\pm\sqrt{\hbar^2\lambda^2(k_x^2+
k_y^2-k_0^2)^2+\hbar^2v^2k_z^2}$ defines a band-degeneracy line
$k_{x}^{2}+k_{y}^{2}=k_{0}^{2}$, that is a nodal loop with a radius $k_0$
in the reciprocal space. The energy scale near the band crossing is of most interest,
where the Hamiltonian \eqref{1} can be linearized and parameterized to \cite{chen2019prl2}
	\begin{equation}\label{2}
	H=\hbar v_{0} \kappa (\cos{\varphi} \tau_{x}+\sin{\varphi} \tau_{y}),
	\end{equation}
through the substitution $k_{x}=(k_{0}+\kappa \cos{\varphi}) \cos{\theta}$,
$k_{y}=(k_{0}+\kappa \cos{\varphi}) \sin{\theta}$,
$k_{z}=\kappa \sin{\varphi}/\alpha $ with $v_{0}=2\lambda k_{0}$
and $\alpha =v/v_{0}$ the ratio between the velocity along the $z$ direction and
that in the ${x}$-${y}$ plane. The parameters
$\kappa,\theta,\varphi$ are the coordinates defined
on the torus as shown in Fig. \ref{fig1}(a).

In the presence of spin-orbit
coupling, a mass term should be added as \cite{zhang2019jpcc}
\begin{equation}\label{3}
\mathcal{H}=H+\Delta\tau_{z}\sigma_z,
\end{equation}
with its sign depending on that of the spin $\sigma_z$. Accordingly, the energy spectrum
becomes $\epsilon^\pm_{\bm{k}}(\kappa)=\pm \sqrt{(\hbar v_{0}\kappa)^{2}+\Delta^{2}}$,
in which a gap of $2\Delta$ is induced along the nodal loop.
We focus on the regime that the effect due to the Zeeman coupling between
the magnetic field and the spin is negligibly small.
Then it can be shown that the main results do not rely on the
spin or equivalently, the sign of the mass, so that
we set $\sigma_z=1$ in the following.

Without loss of generality, we consider a positive Fermi energy,
such that only the electrons in the conduction band contributes
to the transport and the wave function is found to be
$|u_{\bm{k}}\rangle=(\kappa e^{-i \varphi}/\sqrt{\kappa^2+\gamma^2},
\gamma/\sqrt{\kappa^2+\gamma^2})
$ with $\gamma=\sqrt{m^2+\kappa^2}-m$ and $m=\Delta/(\hbar v_0)$.
The Berry curvature defined by $\bm{\Omega}_{\bm{k}}
=i\langle \nabla_{\bm{k}}u_{\bm{k}}|\times|\nabla_{\bm{k}}u_{\bm{k}}\rangle$
has only a nonzero value along the toroidal direction in the parametric coordinates as
\begin{equation}\label{5}
\bm{\Omega}_{\bm{k}}=-\frac{m}{2(\kappa^{2}+m^{2})^{\frac{3}{2}}}\hat{\bm{e}}_{\theta},
\end{equation}
which reduces to a Berry flux density as $m\rightarrow0$.
The Berry curvature induces a self-rotation of the electron wave
packet which gives rise to a finite
OMM \cite{chang1996prb,xiao2010rmp}. In the present two band
system, the OMM is related to the Berry curvature
through $\bm{\mathcal{M}}_{\bm{k}}=\frac{e}{2\hbar}
\bm{\Omega_{\bm{k}}}(\epsilon^+_{\kappa}-\epsilon^-_{\kappa})$, yielding
\begin{equation}\label{6}
\bm{\mathcal{M}}_{\bm{k}}=-\frac{m e v_{0}}{2(\kappa^2+m^2)}\hat{\bm{e}}_{\theta},
\end{equation}
which is also a vector field along the toroidal direction.

One can infer that as an external magnetic field
is imposed, only its components in the $x$-$y$ plane
couple to the OMM.
Without loss of generality,
we consider a magnetic field $\bm{B}=B\hat{\bm{x}}$ in the $x$-direction,
which gives rise to a coupling energy as
\begin{equation}
\delta\epsilon_{\bm{k}}(\kappa,\theta)=-\bm{\mathcal{M}}_{\bm{k}}\cdot{\bm{B}}
=-(m e v_{0}B)\sin\theta/[2(\kappa^2+m^2)].
\end{equation}
The whole energy of the conduction band becomes
\begin{equation}
\tilde{\epsilon}_{\bm{k}}(\kappa,\theta)=\epsilon^+_{\bm{k}}(\kappa)+\delta\epsilon_{\bm{k}}(\kappa,\theta),
\end{equation}
which now depends on the toroidal angle $\theta$.
A direct result of this anisotropy is a deformation of the Fermi
torus, as shown in Figs. \ref{fig1}(c)-\ref{fig1}(e).
We will demonstrate in the following that such a deformation
is the main scenario of the magnetotransport.

\section{Semiclassical theory}\label{III}
In this section, we solve the transport problem
in the presence of impurity scattering
using the semiclassical Boltzmann formalism.
We assume that the magnetic field satisfies $B\ll \epsilon_F/(v_0^2e\tau)$
with $\epsilon_F$ the Fermi energy and $\tau$ the elastic scattering time,
which means that the Landau level spacing $\hbar v_0^2 eB/\epsilon_F$ near
the Fermi energy
is much less than the energy uncertainty
$\hbar/\tau$ due to scattering
and thus the effects of the
Landau quantization can be ignored.
The semiclassical equations of motion are given by \cite{sundaram1999prb,xiao2010rmp}
\begin{equation}\label{7}
\dot{\bm{r}}=\frac{1}{\hbar}\nabla_{\bm{k}}\tilde{\epsilon}_{\bm{k}}-\dot{\bm{k}}\times{\bm{\Omega}_{\bm{k}}},\qquad \dot{\bm{k}}=-\frac{e}{\hbar}(\bm{E}+\dot{\bm{r}}\times{\bm{B}}),
	\end{equation}	
where $\bm{r}$ and $\bm{k}$ are the central position and momentum
of the wave packet, respectively. $-e$ is the electron charge, and $\bm{E}$ and $\bm{B}$
are the external electric and magnetic fields.
We focus on the linear-response regime ($\bm{E}=0$)
and solving Eq. (\ref{7}) gives an effective velocity as
\begin{equation}\label{8}
\dot{\bm{r}}=[\tilde{\bm{v}}_{\bm{k}}+(e/\hbar)\bm{B}(\tilde{\bm{v}}_{\bm{k}}\cdot{\bm{\Omega}_{\bm{k}}})]/D_{\bm{k}},
\end{equation}
where $D_{\boldsymbol{k}}^{-1}$ is the correction to the density of states \cite{xiao2005prl} with
\begin{equation}\label{9}
D_{\boldsymbol{k}}=1+\frac{e}{\hbar} \boldsymbol{B}\cdot{\boldsymbol{\Omega_{k}}},
\end{equation}
and $\tilde{\bm{v}}_{\bm{k}}=(1/\hbar)\nabla_{\bm{k}}\tilde{\epsilon}_{\bm{k}}$ yields
\begin{equation}
\tilde{\bm{v}}_{\bm{k}}=\bm{v_{k}}-\frac{1}{\hbar}\nabla_{\bm{k}}(\bm{\mathcal{M}_{k}}\cdot{\bm{B}}).
\end{equation}
One can see that apart from the pure effect of the
Berry curvature on the velocity in Eq. \eqref{8},
the OMM introduces a correction of the velocity as
$\bm{v}_{\bm{k}}\rightarrow\tilde{\bm{v}}_{\bm{k}}$,
which is essential for the MR in the massive nodal-line semimetals.
Similarly, we have
\begin{equation}\label{12}
\dot{\bm{k}}=-\frac{e}{\hbar D_{\bm{k}}}\Big[\bm{E}+ \tilde{\bm{v}}_{\bm{k}}\times{\bm{B}}+\frac{e}{\hbar}(\bm{B}\cdot{\bm{E}})\bm{\Omega_{k}}\Big].
\end{equation}

The steady Boltzmann equation is interpreted as
\begin{equation} \label{10}
\dot{\bm{k}}\cdot{\nabla_{\bm{k}}} {f}=-\frac{{f}-{f}_{0}}{\tau},
\end{equation}
where a uniform condition and the relaxation time approximation are adopted.
The function ${f}$ is the distribution function and
${f}_ {0}=1/[e^{(\tilde{\epsilon}_{\bm{k}}-\epsilon_{F})/k_BT}+1]$ is
the Fermi-Dirac equilibrium distribution.
Substituting Eq. \eqref{12} into  Eq. \eqref{10} and
keeping the terms to the first order of $\bm{E}$ on both sides yields
\begin{equation}\label{13}
{f}_{1}=\frac{e\tau}{D_{\bm{k}}}\Big[\bm{E}+\frac{e}{\hbar} (\bm{E}\cdot{\bm{B}})\bm{\Omega_{k}}\Big]\cdot\tilde{\bm{v}}_{\bm{k}} \frac{\partial{{f}_{0}}}{\partial{\tilde{\epsilon}}}.
\end{equation}
where ${f}_{1} = {f}-{f}_{0}$ is the first-order
deviation of the distribution from the equilibrium.
The current density is solved by
\begin{equation}\label{11}
\bm{J}=-e \int \frac{d\bm{k}}{(2\pi)^3} {f}_{1} D_{\bm{k}}\dot{\bm{r}},
\end{equation}
and the longitudinal conductivity defined by $J_\mu=\sigma_{\mu}E_\mu\ \ (\mu=x,y,z)$
takes the form of \cite{dai2017prl}
	\begin{equation}\label{14}
	\sigma^{\mu}=-\int \frac{d^{3} \bm{k}}{(2\pi)^3} \frac{e^{2}\tau}{D_{\bm{k}}} (\tilde{v}_{\bm{k}}^{\mu}+\frac{e}{\hbar} B^{\mu}\tilde{\bm{v}}_{\bm{k}}\cdot\bm{\Omega}_{\bm{k}})^{2} \frac{\partial{{f}_{0}}}{\partial{\tilde{\epsilon}}}
	\end{equation}
	
\section{Manetotransport properties}\label{IV}
First, we note that a magnetic field $\bm{B}=B\hat{\bm{z}}$ in the $z$-direction
has no coupling to the OMM, i.e., $\bm{\mathcal{M}}_{\bm{k}}\cdot{\boldsymbol{B}}=0$,
which does not result in any deformation of the Fermi torus.
Accordingly, we have $\tilde{\bm{v}}_{\bm{k}}=\bm{v}_{\bm{k}}$
and $\tilde{\bm{v}}_{\bm{k}}\cdot\bm{\Omega}_{\bm{k}}=0$ in Eq. (\ref{14}),
which indicates that no $\bm{B}$-dependence of the conductivity or MR appears.
Hereafter, we consider the external magnetic field applied in the $x$-direction,
so that a coupling between the OMM and the magnetic field can be achieved.
It induces a deformation of the Fermi torus
which increases with $B$ and the gap $\Delta$ induced by the
spin-orbit coupling. Moreover, the conductivity in Eq. \eqref{14}
depends on $\bm{B}$ in general, indicating a MR induced
by the deformation of the Fermi torus.

\subsection{Symmetry analysis}
Before we go to further details, let us first
perform a symmetry analysis on the conductivity $\sigma(B,m)$.
Note that Eq. (\ref{14}) is invariant as $B$ and $m$ change their signs simultaneously,
that is
\begin{equation}\label{15}
\sigma^{\mu}(B,m)= \sigma^{\mu} (-B,-m).
\end{equation}
Additionally, the nodal-line semimetal possesses
the rotation symmetry about the $z$-axis,
which means that as we invert both $\bm{B}$ and
$\bm{E}$ (equivalent to a rotation of
the fields by $\pi$), the conductivity remains the same,
that is
\begin{equation}\label{16}
\sigma^{\mu}(B,m)=\sigma^{-\mu}(-B,m).
\end{equation}
Finally, the following relation holds,
\begin{equation}\label{17}
\sigma^{\mu}(B,m)=\sigma^{-\mu}(B,m),
\end{equation}
which ensures that there is no spontaneous
current flowing as $\bm{E}=0$.
Combining Eqs. \eqref{15}, \eqref{16} and \eqref{17}
leads to
\begin{equation}\label{18}
\begin{split}
\sigma^{\mu}(B,m)=\sigma^{\mu}(B,-m),\\
\sigma^{\mu}(B,m)= \sigma^{\mu}(-B,m),
\end{split}
\end{equation}
which shows that $\sigma^\mu(B,m)$ is an even function of both $B$ and $m$
and confirms that the MR is independent of the sign of the mass as mentioned before.
It is convenient to separate the conductivity to two terms as
$\sigma^\mu(B,m)=\sigma^\mu_0+\delta\sigma^\mu(B,m)$,
with $\sigma^\mu_0$ the Drude conductivity for $B=0$
and $\delta\sigma^\mu(B,m)$ being the magnetoconductivity, which
scales as $\propto B^2m^2$ to the lowest order of $B$ and $m$
according to the symmetry analysis above.

\subsection{Slight Fermi torus deformation and positive MR}
In order to get some analytic results,
we first solve the conductivity \eqref{14}
in the weak field limit $B\ll\epsilon_F^2/(\hbar e v_0^2)$,
i.e., the Landau level spacing is much smaller than
the Fermi energy.
This condition is equivalent to $\kappa_F \ell_B\gg1$,
with $\kappa_F=\epsilon_F/(\hbar v_0)$ the Fermi wave vector
and $\ell_B=\sqrt{\hbar/(eB)}$ the magnetic length.
Moreover, we also assume that $\kappa_F\gg m$,
such that the spin-orbit coupling can be considered as
a perturbation. The zero-temperature conductivity
calculated by Eq. \eqref{14} is obtained as (see the Appendix \ref{A} for details)
\begin{equation}\label{cond}
\begin{split}
\sigma_0^x&=\sigma_0^y=\sigma_0^z/{2\alpha^2}=\frac{e^2k_0\epsilon_F\tau}{8\pi\alpha\hbar^2},\\
\delta\sigma^x&=\delta\sigma^y/3=\delta\sigma^z/{4\alpha^2}=-\frac{3\alpha e^4v_0^6\hbar^2k_0\tau}{32\pi \epsilon_F^5}B^2m^2,
\end{split}
\end{equation}
which is consistent with our symmetry analysis that
$\delta\sigma^\mu \propto B^2m^2$.
The magnetic conductivities in three directions
differ from each other by noting that the deformation of the
Fermi torus breaks the rotational symmetry in the $x$-$y$
plane.

Here, the coupling between the OMM and the magnetic field
and the resultant deformation of the Fermi surface
play a decisive role for the magnetoconductivity;
The pure Berry curvature effect cannot lead to this result.
This is in stark contrast to the situations for the Weyl semimetal \cite{son2013prb}
and the topological insulator \cite{dai2017prl},
where the finite Berry curvature leads to a negative MR.
The results in Eq. \eqref{cond} indicate a \emph{positive} longitudinal MR
in the weak field limit
defined by
\begin{equation}\label{22}
\text{MR}_{\mu}(B)=\frac{1/\sigma^{\mu}(B)-1/\sigma^{\mu}(0)}{1/\sigma^{\mu}(0)}.
\end{equation}
Therefore, the MR ($\propto B^2m^2$) induced by the Berry curvature
through the OMM effect in the massive
nodal-line semimetal is opposite in sign to that of the Weyl
semimetal (in the absence of inter-node scattering) \cite{son2013prb,knoll2020prb}
and the topological insulator \cite{dai2017prl}.
Note that the OMM induced deformation of the Fermi surface
also introduces a positive correction to the MR in the Weyl semimetal \cite{knoll2020prb},
which is in agreement with our results.

\subsection{Lifshitz transition and sign reversal of MR}
	
\begin{figure}
\centering
\includegraphics[width=0.4\textwidth]{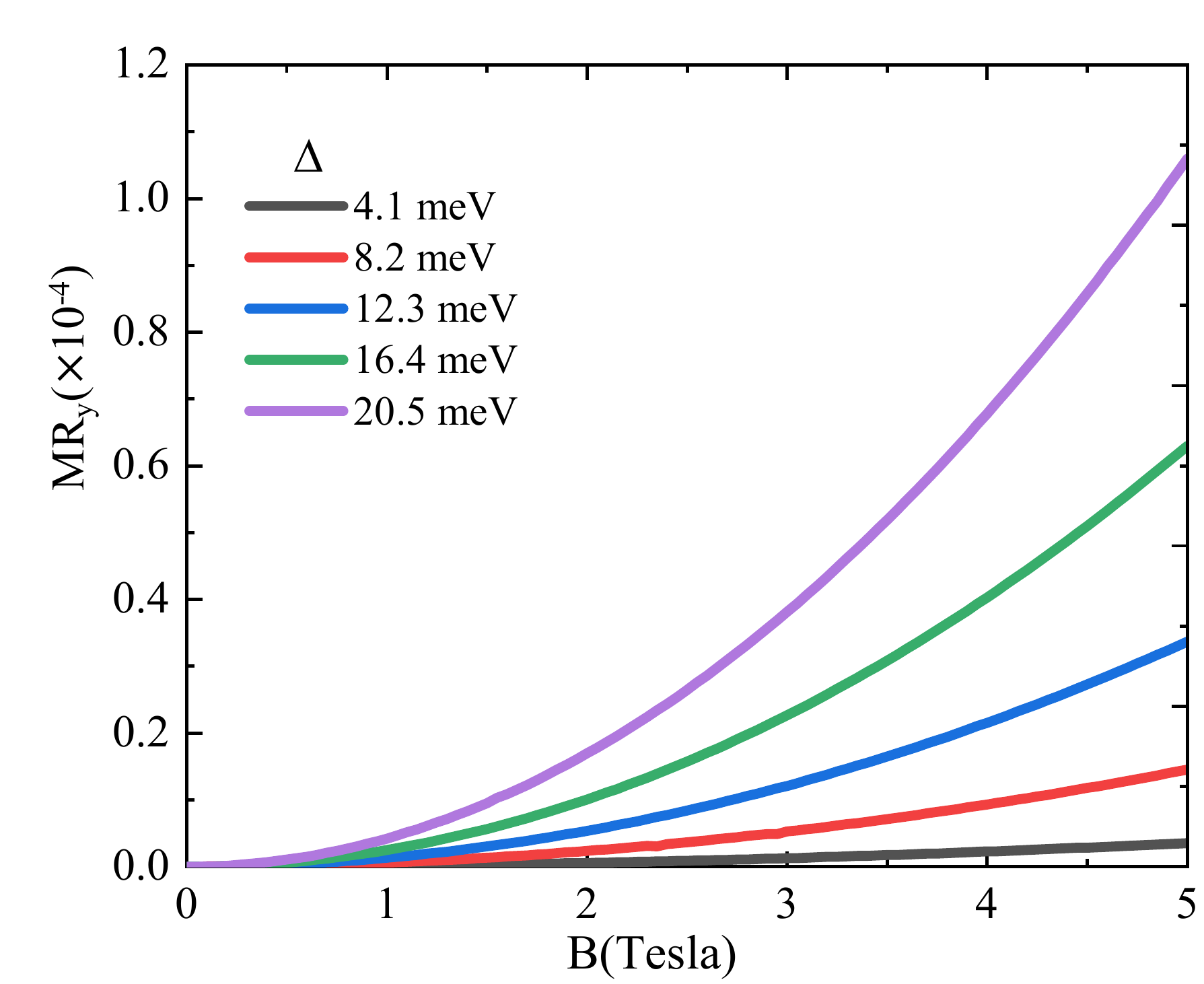}
\caption{MR as a function of magnetic field $B$ for
different gap $\Delta$ in the regime of slight Fermi surface deformation.
The relevant parameters are $\epsilon_{F}=41$ meV,
$T=5$ K, $k_{0}=10$ nm$^{-1}$, $\tau=10^{-14}$ s and $v_{0}=10^{5}$ m/s.}
\label{fig2}
\end{figure}

\begin{figure*}
\includegraphics[width=0.8\textwidth]{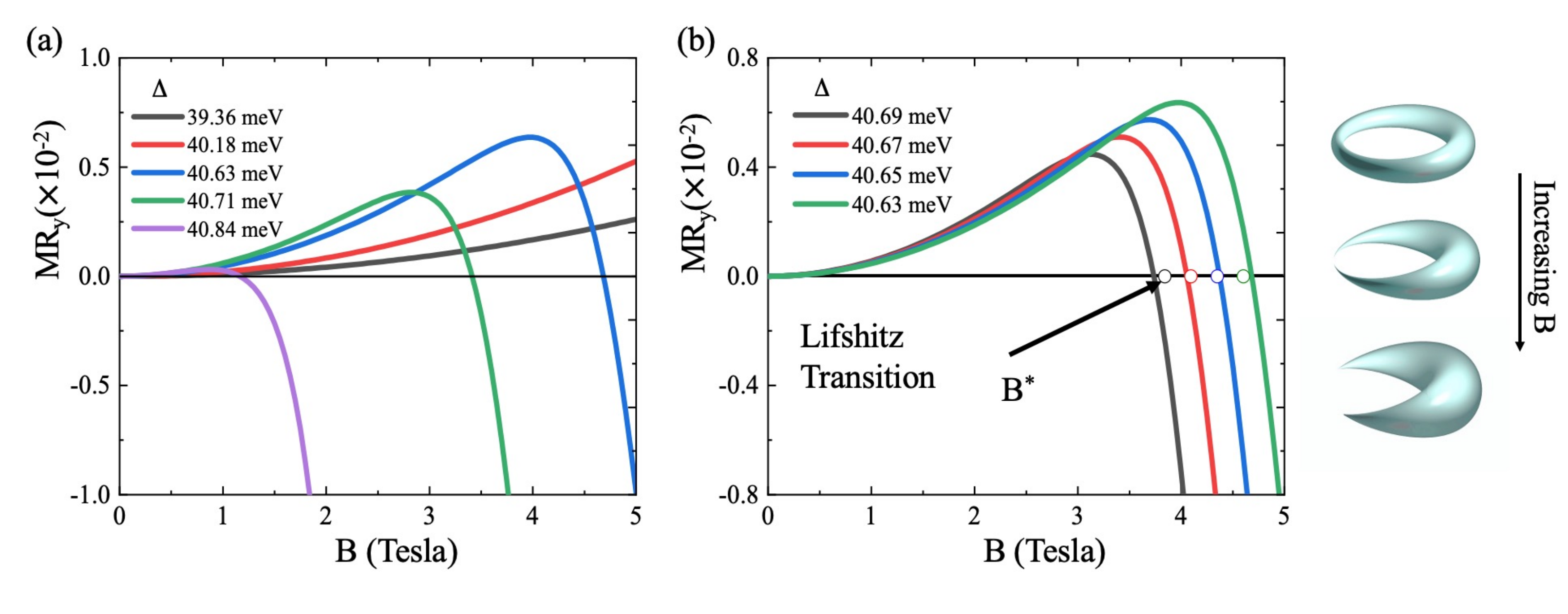}
\caption{(a) MR as a function of $B$ for
different gap $\Delta$ with and without sign reversal.
(b) Correspondence
between the sign reversal point and the critical field $B^*$ (open circles) for the Lifshitz phase transition.
The right panel of (b) illustrates the Lifshitz transition induced by the magnetic field,
which can be compared with the MR results.
Other parameters are the same as those in Fig .\ref{fig2}.}
\label{fig3}
\end{figure*}

In the previous section, our analytical results show that a
slight deformation of the Fermi torus yields a positive
MR which exhibits the $B^2$ dependence of the magnetic field.
In this section, we evaluate the conductivity in Eq. \eqref{14}
numerically for more general cases.
First, the numerical results for the limit $\kappa_{F}\gg m$ shown
in Fig. \ref{fig2} confirm the $B^2$ scaling of the positive MR.
A stronger deformation of the Fermi torus occurs as either
$\bm{B}$ or the ratio $m/\kappa_F$ increases, which
yields a stronger coupling between $\bm{B}$
and the OMM. Interestingly, the MR does not obey
a monotonic dependence on $B$,
which can be seen in Fig. \ref{fig3}(a).
The MR first goes up to its maximum
and then drops rapidly as $B$ increases;
Dramatically, the MR changes its sign from
positive to negative as $B$ increases further.

The sign reversal of the MR induced by the
Berry curvature cannot be found in
other topological systems such as Weyl semimetals \cite{son2013prb}
and topological insulators \cite{dai2017prl}, which indicates
a possible new physical scenario of the MR in the massive
nodal-line semimetal. Note that a key difference between
the nodal-line semimetal and other topological systems
lies in its nontrivial Fermi surface topology.
In contrast to a conventional Fermi sphere,
the Fermi torus may undergo a topological Lifshitz transition
at some critical point from genus one to zero [cf. Fig. \ref{fig1}].
Such abrupt change of the Fermi surface
can lead to anomalous physical properties \cite{lifshitz1960JETP,nishimura2019prl,
chu1970prb, norman2010prb,
sandeman2003prl,watlington1977prb,yamaji2007jpc,yelland2011np,liu2010np},
which provides a possible explanation for the
sign reversal of the MR. In the following, we prove
that it is the Lifshitz transition
that inverts the sign of the MR.

Next, we focus on the limit $\kappa_F\ll m$
which generates a thin Fermi torus with the Lifshitz transition
taking place with an accessible field strength.
We mark the critical
field $B^*$ for the Lifshitz
transition points by open circles in Fig. \ref{fig3}(b).
One can see a good coincidence between $B^*$
and the sign reversal points of the MR.
The numerical results are obtained by setting the
temperature to $T=0. 25$ K rather than absolute zero
(due to the computational capability).
We have confirmed the tendency that the match between critical $B^*$
and the sign reversal points gets
better as $T$ decreases,
indicating a strong connection between
the sign change of the MR and the topological
Lifshitz transition of the Fermi torus.

Moreover, the scenario of topological Lifshitz transition induced
sign reversal of the MR can be
proved strictly in the limit $\kappa_F\ll m$, where
the energy can be expressed as $\epsilon_{F}\simeq\hbar v_{0}(m+\eta)$ with
the parameter $\eta\ll m$. The critical field
for the Lifshitz transition can be obtained as $B^{*}=2 \hbar m \eta/e$.
By expanding the expression of the conductance in Eq. (\ref{14})
to the linear order of $\eta$, we obtain $\sigma^\mu(B^{*})=\sigma^\mu(0)$,
which means that the Lifshitz transition coincides
exactly the sign reversal point of the longitudinal MR
in all three directions [see Appendix \ref{proof}
for details].

\begin{figure*}
\includegraphics[width=0.8\textwidth]{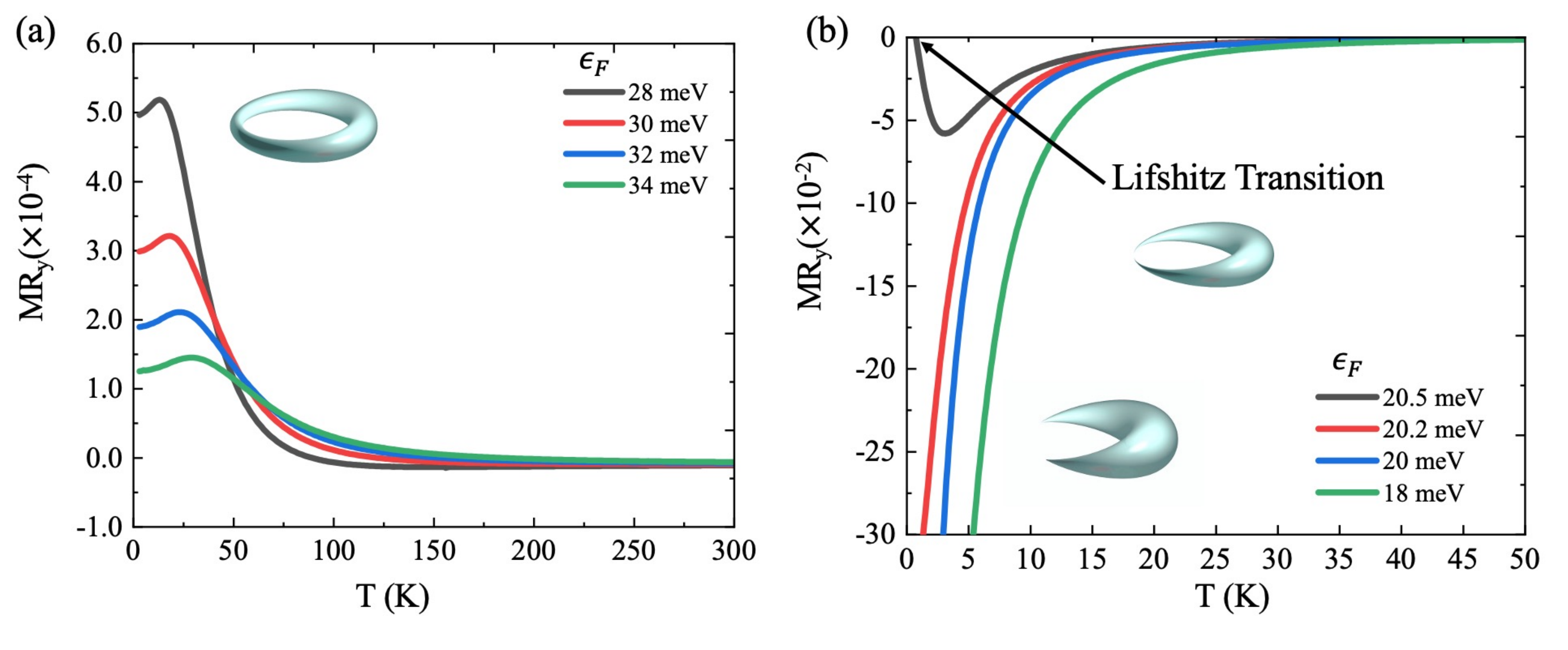}
\caption{Temperature dependence of MR for different Fermi energies $\epsilon_F$ (a) without
and (b) with Lifshitz transition illustrated by the inset.
The parameters are $\Delta=20$ meV, $B=3$ T and the others are the same as those in Fig. \ref{fig2}.}
\label{fig4}
\end{figure*}

\subsection{Finite temperatures}
Next, we investigate the effect of finite temperature on the MR
in Fig. \ref{fig4}. We choose the magnetic field $B$ and the gap $\Delta$ to be fixed
so that the deformation of the Fermi surface is determined by the
energy. As $\epsilon_F>20.5$ meV, the critical value for the
Lifshitz transition defined at zero temperature,
the genus of the Fermi surface is one, corresponding to
a slight deformation; see the inset of Fig. \ref{fig4}(a).
In this regime, the positive MR first increases with temperature
and then undergoes a rapid decrease to zero, exhibiting a non-monotonic
behavior.
For $\epsilon_F<20.5$ meV, Lifshitz transition occurs and
the genus of the Fermi surface becomes zero; see the inset of Fig. \ref{fig4}(b).
Correspondingly, the
MR is negative and its magnitude decreases monotonically to zero
as the temperature rises; see Fig. \ref{fig4}(b).

One can see that in the two regimes with
genus equal to one or zero, the temperature
cannot invert the sign of the MR.
This indicates that the sign of the MR provides a
robust signature of the topology of the Fermi surface
against finite temperature effect. Moreover,
the different temperature dependence of the MR in the two
regimes can serve as an additional manifestation for the identification
of the Fermi surface topology.
For the low temperature, a smaller magnitude of the MR corresponds to
a larger Fermi energy in Figs. \ref{fig4}(a) and \ref{fig4}(b). This stems from that
for a higher energy, the coupling between the OMM
and the magnetic field becomes weaker,
which has the same effect as the
reduction of the gap $\Delta$ as shown in Fig. \ref{fig2}.
For sufficiently high temperature
$k_BT\sim \epsilon_F-\Delta$, the Fermi surface is no longer
well defined so that the MR induced by its deformation quenches.
Finally, at the critical point of Lifshitz transition $\epsilon_F\simeq20.5$ meV,
MR equals zero for both $T=0$ and $T\gg(\epsilon_F-\Delta)/k_B$ and possesses
a negative value in between.

\section{Discussions}\label{V}
It is important to compare our results to other
scenarios of the MR and their manifestation in
the nodal-line semimetals.
The existing transport experiments on the nodal-line semimetals
have reported a variety of transport properties \cite{bannies2021prb,chen2020prb,
an2019prb,guo2019aem,chen2021cpl,laha2020prb,sasmal2020jpcm,li2018sb},
such as large positive MR \cite{chen2020prb,an2019prb,laha2020prb,chen2021cpl,guo2019aem,bannies2021prb}
or cusp-like
behavior of MR in low magnetic field \cite{sasmal2020jpcm,laha2020prb}.
Different mechanisms have been adopted to explain these results,
such as the electron-hole compensation \cite{bannies2021prb,chen2020prb,chen2021cpl}
and weak localization and anti-localization \cite{sasmal2020jpcm,laha2020prb,guo2019aem}.
The former results in a positive MR with a $B^2$ dependence,
no much different from that in a conventional metal \cite{kittel1996};
The latter may lead to either weak localization or
anti-localization with $\sqrt{B}$ or $\ln B$ scaling
which is determined by the type of the impurity
the resultant diffusion dimensionality \cite{chen2019prl}.

The OMM induced MR predicted in our work can be
discriminated from both mechanisms above. First,
the MR exhibits a non-monotonic
$B$ dependence for a strong deformation of the
Fermi torus which can even invert its sign as $B$ approaches $B^*$ [cf. Fig, \ref{fig3}].
Such a phenomenon has not been discovered previously,
which can be explained neither by the electron-hole
compensation nor the weak
(anti-)localization effect.
It can serve as a unique transport property
of the topological Lifshitz transition
induced by the coupling between OMM and magnetic field.
Second, the MR is positive and proportional to $B^2$ for
the slight deformation of the Fermi torus when $B\ll B^*$
[cf. Fig. \ref{fig2}], which
differs in sign from the weak localization
and in scaling law from the weak anti-localization scenario \cite{sasmal2020jpcm,laha2020prb,guo2019aem}.
The mechanism of electron-hole compensation results in
the same sign and scaling, but it gives
an MR whose magnitude is several orders larger than
that in our results \cite{bannies2021prb,chen2020prb,chen2021cpl}.
Third, the distinctive temperature dependence of the MR
in the two regimes with genus one and
zero [cf. Fig,\ \ref{fig4}] further helps
to discriminate the current mechanism from others.

\section{Conclusions }\label{VI}
To conclude, we predict a novel mechanism for the
magnetotransport in nodal-line semimetals
which stems from the Fermi surface deformation
induced by the coupling between the
OMM and magnetic field. Specifically,
a small deformation results in a positive MR
with a quadratic dependence of $B$; Most interestingly,
the strong distortion causes the topological
Lifshitz transition of the Fermi torus that is
in coincidence with a sign reversal of the MR.
The key ingredient of our mechanism is
the spin-orbit coupling induced
mass term which exists in many nodal-line
semimetal candidates \cite{zhang2019jpcc,chen2019prl}.
Large spin-orbit coupling
and low Fermi energy are favorable for the realization of
strong Fermi surface deformation
and Lifshitz transition.
Our work uncovers a novel scenario of the
MR induced solely by the Fermi surface deformation,
which also opens an avenue for studying
topological Lifshitz transition by
magnetotransport.

\begin{acknowledgments}
We thank Jing Luo for assistance on the
figures. This work was supported by the National
Natural Science Foundation of
China under Grant No. 12074172 (W.C.), No. 11674160 and
No. 11974168 (L.S.), the startup
grant at Nanjing University (W.C.), the State
Key Program for Basic Researches of China
under Grants No. 2017YFA0303203 (D.Y.X.)
and the Excellent Programme at Nanjing University.
\end{acknowledgments}

\appendix
\section{Magnetoconductivity with slight deformation of the Fermi torus}\label{A}
Here, we provide a detailed derivation of the magnetoconductivity
as the deformation of the Fermi torus is weak.
We first solve the conductivity with $\alpha=1$
and then generalize the results to $\alpha\neq1$.
The expression of the conductivity \eqref{14}
contains various terms.
The velocity components in the cartesian coordinates are
\begin{equation}
\begin{split}
\tilde{v}_{x}&=-\tilde{v}_{\theta}\sin{\theta}+\tilde{v}_{\kappa} \cos{\theta}\cos{\varphi},\\
\tilde{v}_{y}&=\tilde{v}_{\theta}\cos{\theta}+\tilde{v}_{\kappa}\sin{\theta}\cos{\varphi},\\
\tilde{v}_z&=\tilde{v}_\kappa\sin\varphi,
\end{split}
\end{equation}
with
\begin{equation}\label{33}
\begin{split}
&\tilde{v}_{k}=v_{0}(\frac{\kappa}{\sqrt{\kappa^2+m^2}}+\frac{B e \kappa m \sin{\theta}}{\hbar (\kappa^{2}+m^{2})^{2}})\\
& \tilde{v}_{\theta}=-\frac{Bemv_{0} \cos{\theta}}{2\hbar (\kappa^{2}+m^{2})(k_{0}+\kappa \cos{\varphi})}\\
& \tilde{v}_{\varphi}=0\\
\end{split}
\end{equation}
We focus on the weak field limit $B\ll\epsilon_F^2/(\hbar e v_0^2)$
and $m/\kappa_F\ll1$ so that $\epsilon_{\bm{k}}\simeq \hbar v_0\kappa$ and
\begin{equation}\label{21}
\begin{split}
&\tilde{v}_{\kappa}\simeq v_{0}\Big(1+\frac{m\sin\theta}{\ell_B^2\kappa^3}\Big),\\
& \tilde{v}_{\theta}\simeq -v_0\frac{m\cos\theta}{2\ell_B^2\kappa^2k_0},\\
& \tilde{v}_{\varphi}=0.\\
\end{split}
\end{equation}
The correction to the density of states reduces to
\begin{equation}
\frac{1}{D_{\bm{k}}}\simeq1-\frac{m\sin\theta}{2\ell_B^2\kappa^3}.
\end{equation}
Note that the distribution function $f_0$ in Eq. \eqref{14}
is a function of $\tilde{\epsilon}_{\bm{k}}$ instead of
$\epsilon_{\bm{k}}$. In the aforementioned limit,
we have
\begin{equation}
\delta\epsilon_{\bm{k}}\simeq-\hbar^3mv_0^3\sin\theta/(2\ell_B^2\epsilon^2),
\end{equation}
and accordingly
\begin{equation}
\begin{split}
\frac{\partial f_0}{\partial \tilde{\epsilon}}&=\frac{\partial f_0(\epsilon+\delta\epsilon)}{\partial\epsilon}\frac{\partial\epsilon}{\partial \tilde{\epsilon}}\\
&\simeq\frac{\partial [f_0(\epsilon)+f_0'(\epsilon)\delta\epsilon]}{\partial\epsilon}\Big(1-\frac{\partial\delta\epsilon}{\partial\epsilon}\Big)\\
&=(f_0'+\delta\epsilon'f_0'+\delta\epsilon f_0^{''})(1-\delta\epsilon'),
\end{split}
\end{equation}
which further reduces to
\begin{equation}
\frac{\partial f_0(\tilde{\epsilon})}{\partial \tilde{\epsilon}}\simeq \frac{\partial f_0(\epsilon)}{\partial\epsilon}+\delta\epsilon\frac{\partial^2 f_0(\epsilon)}{\partial\epsilon^2},
\end{equation}
by keeping the terms to the linear order of $\delta\epsilon$.
By inserting all these simplified terms into the integration in Eq. \eqref{14}
and transferring the integral variables to $(\kappa,\theta,\varphi)$,
the conductivity at zero temperature can be obtained as
$\sigma^\mu=\sigma_0^\mu+\delta\sigma^\mu$ with
\begin{equation}
\begin{split}
\sigma_0^x&=\sigma_0^y=\sigma_0^z/2=\frac{e^2k_0\epsilon_F\tau}{8\pi\hbar^2},\\
\delta\sigma^x&=\delta\sigma^y/3=\delta\sigma^z/4=-\frac{3e^4v_0^6\hbar^2k_0\tau}{32\pi \epsilon_F^5}B^2m^2,
\end{split}
\end{equation}
in which the condition $k_0\gg\kappa_F$ is adopted.

The results for $\alpha\neq1$
can be obtained straightforwardly by the
substitutions
$k_z\rightarrow \alpha k_z$ and accordingly,
$\Omega_z(k_z)\rightarrow\Omega_z(\alpha k_z)$,
$\Omega_{x,y}(k_z)\rightarrow\alpha\Omega_{x,y}(\alpha k_z)$,
$\mathcal{M}_z(k_z)\rightarrow\mathcal{M}_z(\alpha k_z)$,
$\mathcal{M}_{x,y}(k_z)\rightarrow\alpha\mathcal{M}_{x,y}(\alpha k_z)$,
which yields
\begin{equation}
\begin{split}
\sigma_0^x&=\sigma_0^y=\sigma_0^z/{2\alpha^2}=\frac{e^2k_0\epsilon_F\tau}{8\pi\alpha\hbar^2},\\
\delta\sigma^x&=\delta\sigma^y/3=\delta\sigma^z/{4\alpha^2}=-\frac{3\alpha e^4v_0^6\hbar^2k_0\tau}{32\pi \epsilon_F^5}B^2m^2,
\end{split}
\end{equation}
which is the Eq. \eqref{cond} in the main text.

\section{Proof of the correspondence between the Lifshitz transition and sign reversal of MR}\label{proof}
In this section, we prove the coincidence between the Lifshitz phase
transition of the Fermi torus and the sign reversal point of the MR
in the limit $\kappa_F\ll m$.
Without loss of generality, we focus on the longitudinal
conductivity in the $y$ direction and express Eq. \eqref{14}
in the parametric coordinates as
\begin{equation}\label{B1}
\begin{split}
\sigma^{y}(B)&=-\int \frac{e^{2}\tau}{(2\pi)^3} \frac{\kappa (k_{0}+\kappa \cos{\varphi})}{D_{\bm{k}}}\\
&\times(\tilde{v}_{\theta}\cos{\theta}+\tilde{v}_{\kappa}\sin{\theta}\cos{\varphi})^2
\frac{\partial{f_{0}}}{\partial{\tilde{\epsilon}}} d\kappa d\theta d\varphi,
\end{split}
\end{equation}
where $\tilde{v}_\theta$ and $\tilde{v}_{\kappa}$ take the general form in Eq. \eqref{33}.
Note that $\tilde{v}_{\theta}\ll \tilde{v}_{\kappa}$ generally holds,
we can neglect $\tilde{v}_\theta\cos\theta$ term in the parentheses.
The Lifshitz transition can be induced by
an accessible magnetic field in the limit $\kappa_F\ll m$,
in which the energy can be expressed as
$\epsilon_{F}= \hbar v_{0}(m+\eta)$ with $\eta \ll m$.
The critical field $B^*$ for the Lifshitz transition
is defined by
\begin{equation}
\tilde{\epsilon}(\kappa=0,\theta=-\pi/2,B=B^*)=\epsilon_F,
\end{equation}
which gives the critical magnetic field
\begin{equation}\label{B5}
B^{*}=2\hbar m\eta/e
\end{equation}
by neglecting $(\kappa/m)^2$ terms.

Next we evaluate the integration in Eq. \eqref{B1}
at the critical point $B^*$ in which the relevant terms
can be simplified to
\begin{equation}
\begin{split}
\frac{\partial f_0}{\partial\tilde{\epsilon}}&\simeq \delta[\hbar v_{0}\kappa^{2}/(2m)-\hbar v_0\eta(\sin{\theta}+1)],\\
\frac{1}{D_{k}}&\simeq(1-\frac{\eta}{m}\sin{\theta}),\\
\tilde{v}_\kappa&=v_{0}\frac{\kappa}{\sqrt{m^{2}+\kappa^{2}}}(1+\frac{2\eta\sin{\theta}}{m})
\end{split}
\end{equation}
in the limit $\kappa\ll m$. By performing the integration
by keeping the term $\eta/m$ to its first order we arrive at
\begin{equation}
\sigma^y(B^*)=\sigma_0^y,
\end{equation}
which means that at the critical field
for the topological Lifshitz transition,
the magnetoconductivity and thus the MR is zero,
corresponding to the sign reversal point of the MR
[cf. Fig. \ref{fig3}(b) in the main text].
The same conclusion of the MR in the $x$ and $z$ directions
can also be proved in a similar way.

%

\end{document}